\documentclass[conf]{new-aiaa}
\usepackage[utf8]{inputenc}

\usepackage{graphicx}
\usepackage{amsmath}
\usepackage[version=4]{mhchem}
\usepackage{siunitx}
\usepackage{longtable,tabularx}
\usepackage{float}
\usepackage{multirow}
\title{Physics-Informed EvolveGCN: Satellite Prediction for Multi Agent Systems}

\author{Timothy Jacob Huber\footnote{Graduate Researcher, Aerospace, Physics, and Space Science, 150 W University Blvd, Melbourne, FL 32901}, Madhur Tiwari PhD. \footnote{Assistant Professor, Aerospace, Physics, and Space Science, 150 W University Blvd, Melbourne, FL 32901} and Camilo A. Riano-Rios PhD. \footnote{Assistant Professor, Aerospace, Physics, and Space Science, 150 W University Blvd, Melbourne, FL 32901} }

\begin{document}

\maketitle

\begin{abstract}
In the rapidly evolving domain of autonomous systems, interaction among agents within a shared environment is both inevitable and essential for enhancing overall system capabilities. A key requirement in such multi-agent systems is the ability of each agent to reliably predict the future positions of its nearest neighbors. Traditionally, graphs and graph theory have served as effective tools for modeling inter agent communication and relationships. While this approach is widely used, the present work proposes a novel method that leverages dynamic graphs in a forward looking manner. Specifically, the employment of EvolveGCN, a dynamic graph convolutional network, to forecast the evolution of inter-agent relationships over time. To improve prediction accuracy and ensure physical plausibility, this research incorporates physics constrained loss functions based on the Clohessy-Wiltshire equations of motion. This integrated approach enhances the reliability of future state estimations in multi-agent scenarios.
\end{abstract}

\section*{Nomenclature}

\begin{longtable}{@{}ll@{}}
$A$ & Adjacency matrix representing connections between agents \\
$D$ & Degree matrix of the graph \\
$G$ & Graph representing agent interactions \\
$H^{(l)}$ & Node feature matrix at layer $l$ in the GCN \\
$n$ & Mean motion of the chief satellite \\
$t$ & Elapsed time \\
$x, y, z$ & Relative position components in LVLH frame \\
$x_0, y_0, z_0$ & Initial relative positions in LVLH frame \\
$v_x, v_y, v_z$ & Relative velocity components in LVLH frame \\
$v_{x0}, v_{y0}, v_{z0}$ & Initial relative velocities \\
\\
CW & Clohessy-Wiltshire \\
GCN & Graph Convolutional Network \\
GRU & Gated Recurrent Unit \\
LVLH & Local Vertical Local Horizontal \\
MPC & Model Predictive Control \\
TGN & Temporal Graph Network \\
ST-GCN & Spatio-Temporal Graph Convolutional Network \\
\\
$\tilde{A}$ & Normalized adjacency matrix with self loops \\
$\tilde{D}$ & Degree matrix corresponding to $\tilde{A}$ \\
\end{longtable}

\section{Introduction}
\lettrine{A}{utonomous} systems are an ever expanding subject in modern research, with new learning based control schemes for aerospace applications being developed daily \cite{Brunke2022,Sheikhsamad2024}. With this rise in autonomous systems comes the need for cooperation between agents in an environment. Not only to avoid collisions and damages but also to improve productivity by having a group of these agents working together for a common goal. This multi-agent way of thinking requires extensive knowledge of the surrounding agents in a swarm. However, it is not feasible to assume that every agent has extensive knowledge of each agent in the system at all times. Thus, a way to estimate the positional data for nearby agents is critical for multi-agent systems to perform objectives like formation flight safely and without collision \cite{Shalaby2021RAL,Wang2023Sensors}.

Many works have used graph theory to keep a mathematical appendix of the agents and their interactions with other agents in the swarm \cite{Chang2025,olfati2007consensus,ren2005survey,Scardovi2009}. It provides a clear framework for understanding spatial and temporal connections, allowing for efficient algorithms to analyze communication patterns, optimize resource allocation, and facilitate coordination in complex systems. Moreover, graphs enable scalable solutions by representing both static and dynamic networks, accommodating changes in agent states, positions, and interactions over time.

Graph representation can be split into two categories, \textit{static} and \textit{dynamic} representation. Static being a graph that has nodes that do not change over time, while dynamic means the nodes are free to move and interact with each other. Graph neural networks (GNNs) are commonly employed to analyze these graph structures and identify patterns \cite{jiang2022multi}. Many works have used static representation for multi-agent systems, meaning the graph representing the agents and their interactions does not alter over time \cite{zhou2021graphneuralnetworksreview}. However, static representation does not hold the potential to show changing systems such as social networks like multi-agent systems. Dynamic or temporal graphs, by contrast, offer a more accurate representation by allowing nodes to move and form new connections over time. Models such as \textit{EvolveGCN} \cite{pareja2020evolvegcn} are specifically designed to handle these evolving graph structures. EvolveGCN is a machine learning model designed to analyze temporal networks. By leveraging graph convolutional layers (GCNs), it enables dynamic nodes to establish new connections as they evolve over time. In a multi-agent system, this capability allows agents to predict and track the trajectories of their observed neighbors, forming a "nodal neighborhood" that facilitates predictive maneuvers to prevent collisions. This is very simple with graph theory due to the innate way of representing graphs as a matrix, meaning agents and connections can be described as a single matrix. This matrix representation allows for fast and simple computations. While most works that use graphs for multi-agent systems incorporate it as a communication graph where nodes represent the agents in the system and the edges between them representing a communication link where data can be passed along for data sharing between agents \cite{doi:10.2514/1.G008280,doi:10.2514/6.2024-0992,ValenciaPalomo2024}. This passing of information in graph theory is called aggregation between nodes. In graph theory, links and nodes can describe many different things and links can describe any relationship between nodes, not just communication \cite{yan2024control}. This is the basis that this research is based on. 

In this research, EvolveGCN is implemented as a means to propagate a dynamic graph into the future. Each node represents an agent in the system and the links between the agents are within sensing range of the agent in question. The graph of agents and connections are passed through EvolveGCN with a physics informed loss to constrain the position and velocity predictions to within a reasonable bound. 

\section{EvolveGCN}
In this research, EvolveGCN serves as the foundation for modeling agent interactions over time. EvolveGCN combines Graph Convolutional Networks (GCNs) with a matrix based Gated Recurrent Unit (GRU) to capture both spatial dependencies and temporal evolution in dynamic graphs \cite{pareja2020evolvegcn}. This architecture allows the model to adapt the graph structure and node representations over time, making it suitable for scenarios where inter agent relationships change continuously, such as in multi-agent robotic systems.

Previous work has applied EvolveGCN to multi-agent systems to estimate the future positions of nearby agents by leveraging both historical position data and future trajectories provided by Model Predictive Control (MPC) \cite{yan2024control}. However, these implementations largely treat EvolveGCN as a black box temporal propagator and do not explicitly incorporate the underlying physical constraints or dynamics of the agents. This limitation affects the model’s fidelity in aerospace contexts, where adherence to dynamics like inertia, velocity constraints, and control inputs is critical.

Alternative dynamic graph propagation models have also been explored in the literature. Spatio-Temporal Graph Convolutional Networks (ST-GCNs), originally developed for human action recognition \cite{yan2018stgcn}, have been adapted for trajectory prediction in crowd navigation and traffic systems \cite{zhao2020stgcn,kim2021stgrnn}. However, ST-GCNs rely on fixed temporal adjacency and are less effective in environments where agent interactions change rapidly and unpredictably. Dynamic Graph Neural Networks (DGNNs), such as the model proposed by the paper \textit{Dynamic graph convolutional networks}. \cite{manessi2020dynamic}, and their extensions for multi-agent path planning \cite{zhou2022dgnn}, offer improved adaptability through learned message passing, but often assume fixed embedding structures or require dense graph connectivity, which may not be feasible in resource constrained robotic systems.

Temporal Graph Networks (TGNs) represent another class of models that use event driven temporal updates and memory modules for evolving graphs \cite{rossi2020temporal}. TGNs have seen applications in multi-agent vehicle interaction modeling \cite{han2022tgn} and UAV swarm coordination using asynchronous spatiotemporal reasoning \cite{li2023tgn}. While powerful, TGNs are typically optimized for irregular event sequences rather than continuous-time state estimation.

EvolveGCN distinguishes itself from these approaches by dynamically updating the GCN parameters themselves over time, enabling more flexible and expressive modeling of temporal changes in both node features and edge structures. In this work, EvolveGCN is adapted not just as a generic temporal model, but as a physics-informed predictor, where estimated future positions of agents must conform to real world movement constraints. This novel integration opens the possibility for more accurate and safer multi-agent coordination.

\section{Physics-Informed EvolveGCN}
Physics informed neural networks are networks that incorporate dynamics into the loss function of a neural network to constrain the output to be more true to the overarching dynamics of a system. This allows for neural networks to be applied to more complex estimation and trajectory following tasks allowing for more complex and true autonomous robotics. Recent work in Physics Informed Neural Networks (PINNs) has demonstrated the benefits of incorporating physical laws particularly through differential constraints such as accelerations into learning objectives for dynamical systems \cite{raissi2019physics,karniadakis2021physics}. These methods have been applied in fields ranging from fluid dynamics to robotic control, offering improved generalization and physical consistency compared to purely data driven models. More recently, such approaches have been extended to structured systems like robotic swarms and aerospace vehicles, where models integrate position, velocity, and acceleration constraints to ensure physically plausible behavior \cite{li2022deepopinn,lu2021learning}.

This research follows that direction by incorporating a physics-informed loss based on the Clohessy-Wiltshire (CW) equations, which model the relative motion of objects in a circular orbit. The model is trained to predict both position and velocity over time, and the loss function incorporates terms to enforce physical plausibility through CW-based constraints. While this method utilizes the Clohessy-Wiltshire equations this methodology can be applied to any overarching dynamics a swarm would need to follow. 

The standard supervised loss is defined as:

\begin{equation}
\mathcal{L}_{\text{data}} = \lambda_p \cdot \| \hat{p}_t - p_t \|^2 + \lambda_v \cdot \| \hat{v}_t - v_t \|^2,
\end{equation}

where $\hat{p}_t$, $\hat{v}_t$ are the predicted position and velocity, $p_t$, $v_t$ are ground truth values, and $\lambda_p$, $\lambda_v$ are tunable weights.

To encourage physically plausible predictions, physics-based regularization is applied. The predicted acceleration is computed by finite difference:

\begin{equation}
\hat{a}_t = \frac{\hat{v}_t - \hat{v}_{t-1}}{\Delta t},
\end{equation}

and the CW-predicted acceleration is computed using the differential form of the CW equations:

\begin{equation}
a^{\text{CW}}_t = 
\begin{bmatrix}
3n^2 \hat{x}_t + 2n \hat{v}_{y,t} \\
-2n \hat{v}_{x,t} \\
-n^2 \hat{z}_t
\end{bmatrix},
\end{equation}

where $n$ is the mean motion of the chief satellite.

The physics loss combines both acceleration and propagated position consistency:

\begin{equation}
\mathcal{L}_{\text{physics}} = \| \hat{a}_t - a^{\text{CW}}_t \|_1 + \| \hat{p}^{\text{CW}}_t - p_t \|_1,
\end{equation}

where $\hat{p}^{\text{CW}}_t$ is the predicted position from the CW equations given $\hat{p}_t$ and $\hat{v}_t$.

The total loss function becomes:

\begin{equation}
\mathcal{L}_{\text{total}} = \mathcal{L}_{\text{data}} + \lambda_{\text{phys}} \cdot \mathcal{L}_{\text{physics}},
\end{equation}

where $\lambda_{\text{phys}}$ ramps up over training to gradually introduce physical constraints.

\vspace{12pt}

The values for velocity and position for each agent is computed using the differential version of the Clohessy-Wilthsire equations shown below: 

\begin{align}
x(t) &= \left(4 - 3\cos(nt)\right)x_0 + \frac{1}{n}\sin(nt) \, v_{x0} + \frac{2}{n}\left(1 - \cos(nt)\right)v_{y0}, \\
y(t) &= 6x_0 \left(\sin(nt) - nt\right) + y_0 - \frac{2}{n}\left(1 - \cos(nt)\right)v_{x0} + \frac{1}{n}\left(4\sin(nt) - 3nt\right)v_{y0}, \\
z(t) &= z_0 \cos(nt) + \frac{1}{n} \sin(nt) \, v_{z0}, \\
v_x(t) &= 3n \sin(nt) x_0 + \cos(nt) v_{x0} + 2 \sin(nt) v_{y0}, \\
v_y(t) &= 6n \left( \cos(nt) - 1 \right)x_0 - 2 \sin(nt) v_{x0} + \left(4 \cos(nt) - 3\right)v_{y0}, \\
v_z(t) &= -n \sin(nt) z_0 + \cos(nt) v_{z0},
\end{align}

\begin{itemize}
    \item $x(t), y(t), z(t)$ are the relative positions in the LVLH (Local Vertical Local Horizontal) frame,
    \item $v_x(t), v_y(t), v_z(t)$ are the corresponding relative velocities,
    \item $x_0, y_0, z_0$ and $v_{x0}, v_{y0}, v_{z0}$ are the initial position and velocity components,
    \item $n$ is the mean motion of the chief satellite (i.e., $\sqrt{\mu/a^3}$),
    \item $t$ is the elapsed time.
\end{itemize}

\vspace{12pt}

This research applies the differential form of the Clohessy-Wiltshire equations to each agent to propagate the dynamics of each agent over a time period. This propagation gives the first component of the node embeddings or stored data the position and velocity at a given timestep. This is coupled with the number of edges or degree of the agent. This degree number shows how many other agents can be sensed and measured from the source agent. When a neighboring agent is within sensing range, a link is established to represent the source agent being able to record measurements from this neighboring agent. This degree number is stored in the embedding as well as the relative position and velocity data of the sensed neighbors from the source agent. The inclusion of the relative motion data allows for the aggregation of data to make more computational sense as without the relative positions and velocities in the data, it would be detrimental to the system for prediction purposes to have connected agents. 

With the node embeddings propagated, the EvolveGCN framework applies graph convolutions across the current graph structure, where each node embedding includes both absolute and relative kinematic data. The graph convolutional layers process this structural information by aggregating messages from neighboring nodes, weighted by the graph's adjacency matrix and the spatial relationships between agents. Following this, the temporal component of EvolveGCN is applied. Unlike traditional GCNs, EvolveGCN uses a matrix-based Gated Recurrent Unit (GRU) to evolve the weights of the graph convolutional layers themselves over time \cite{pareja2020evolvegcn}. This allows the model to not only process changes in node features and connectivity but also to adapt the graph propagation mechanism based on temporal dynamics. As a result, the node embeddings evolve through time in a way that reflects both the physical dynamics of each agent and the evolution of their sensing-based connections.

\begin{align}
H^{(l+1)} &= \tilde{D}^{-1/2} \tilde{A} \tilde{D}^{-1/2} H^{(l)} W^{(l)}, \\
Z_t &= \sigma(W_z z_t + U_z \circ W_{t-1} + B_z), \\
R_t &= \sigma(W_r z_t + U_r \circ W_{t-1} + B_r), \\
\tilde{W}_t &= \tanh(W_h z_t + U_h \circ (R_t \circ W_{t-1}) + B_h), \\
W_t &= (1 - Z_t) \circ W_{t-1} + Z_t \circ \tilde{W}_t,
\end{align}

\begin{itemize}
    \item $H^{(l)}$ is the node feature matrix at layer $l$,
    \item $W^{(l)}$ is the learnable weight matrix for layer $l$,
    \item $\tilde{A} = A + I$ is the adjacency matrix with self loops,
    \item $\tilde{D}$ is the degree matrix corresponding to $\tilde{A}$,
    \item $z_t$ is the graph summary vector at time $t$ (e.g., mean node features),
    \item $W_t$ is the time evolved weight matrix,
    \item $\circ$ denotes element wise (Hadamard) product,
    \item $\sigma$ is the sigmoid activation function.
\end{itemize}

It is with this aggregation technique that separates EvolveGCN from other dynamic graph networks, as the use of a GRU allows for the weights of the GCN to be updated dynamically based on stored temporal data allowing for a more accurate representation of the dynamical representation of the graph and the embeddings of the nodes.

\section{Preliminary Results}
Preliminary results of this physics informed EvolveGCN for multi-agent object tracking and predictions are shown in this section. The physics informed EvolveGCN is trained using the basic model for EvovleGCN that uses the GRU matrix format for a base model. This was fitted with changes to include the physics loss introduced in the section above for weight matrix updates. During training for these results 100 planar trajectories were generated using the Clohessy-Wiltshire equations at different starting orbit distances from 6880 km to 7800 km, with the number of agents varying from 3 to 8 agents. The starting positions and velocities of these agents are randomized within a radius of the chief orbit governed by the random choice of starting orbit altitude. From this an adjacency matrix is generated based on relative distance from neighbors in this synthetic satellite constellation, and links are created based on a distance threshold simulating sensing range of the satellite (if the agent and sense the other agent, create and link between them for data sharing). For the adjacency matrix self connections are used so during the update of each agents node embedding vector, the update takes into account the nodes neighbors and its own past value. During the training each of the trajectories are split into groups of testing validation and training with a 10/20/70 percent split. Each trajectory is its own batch during training, meaning that during training every epoch goes through each of the trajectories independently. The EvolveGCN only looks at 8 time steps in the past per trajectory in a sliding window to predict the next 6 time steps in the future and does this sliding window of history and predictions until the whole trajectory is covered and the batch completes. The average loss of these trajectories in each batch is used to compute the epochs total loss. During training the physics loss weight is kept at maximum of 0.25 to keep it present but not overpowering, this weight ramps up over training from a starting physics weight value of 0.2. A ramp up auto regressive ratio is implement during training in which a percentage of the training trajectories are chosen to use their predictions in the future prediction and history window. Due to these trajectories using auto regression having a inherently higher loss than the other trajectories using ground truth data. In the weighted average of all the trajectories losses, these trajectories are weighted less than the ones using true data. 

Preliminary results from this training is shown below for predicting a 3 satellite constellation position, both using physics in the loss and without.

\begin{figure}[H]
    \centering
    \includegraphics[width=0.85\textwidth]{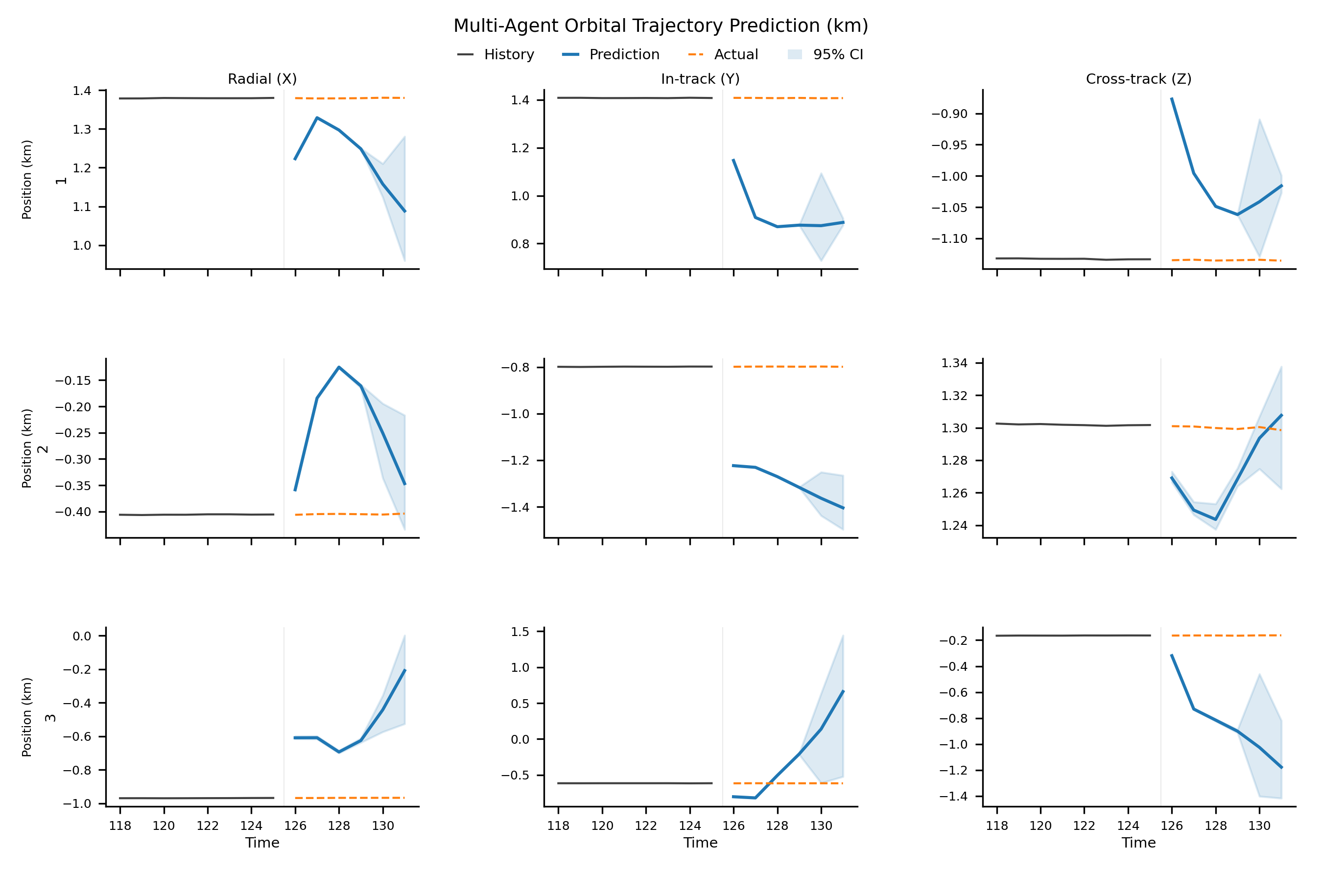}
    \caption{Prediction comparison of a 3 satellite constellation using EvolveGCN without physics loss.}
    \label{fig:3sat_prediction}
\end{figure}

\begin{figure}[H]
    \centering
    \includegraphics[width=0.85\textwidth]{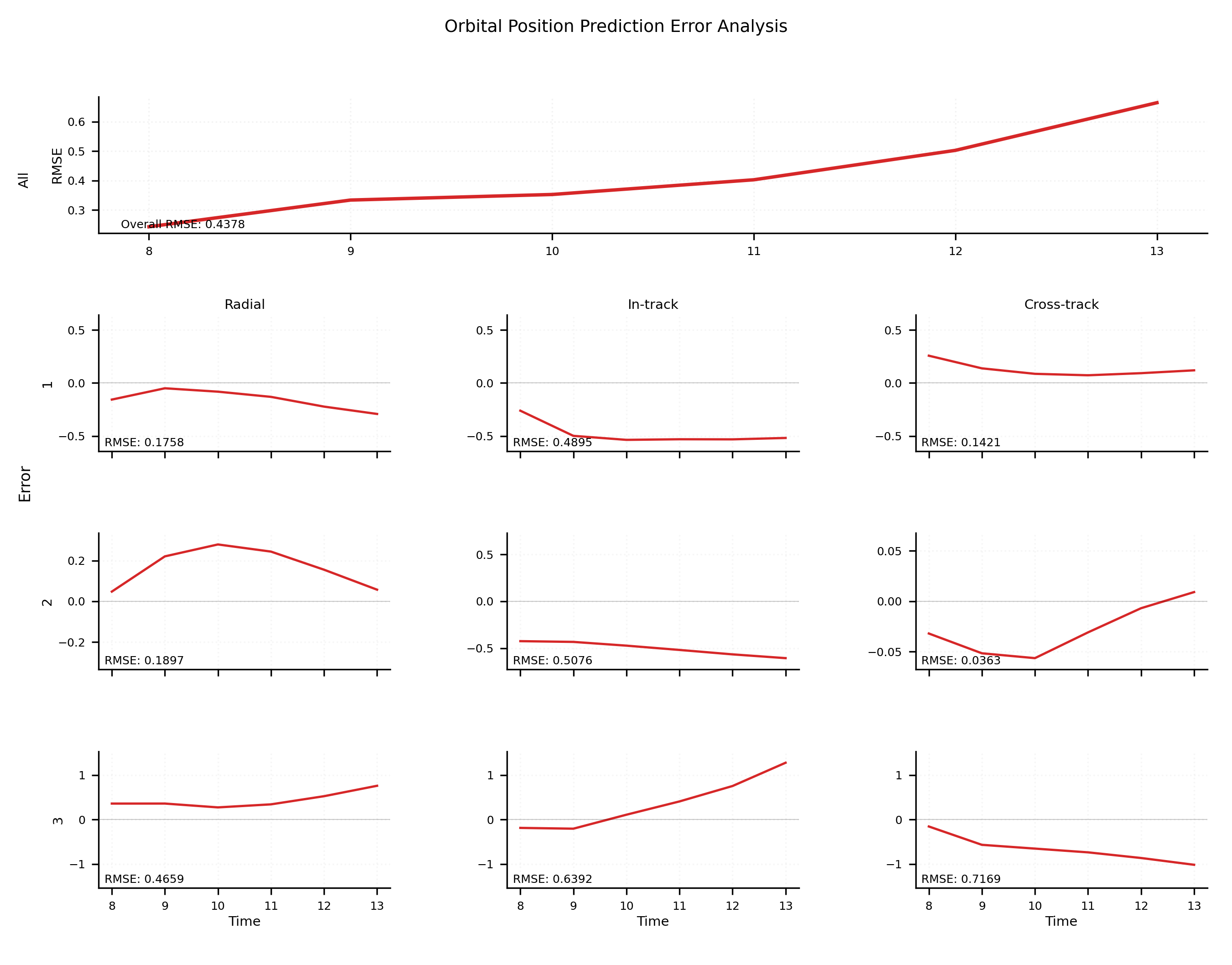}
    \caption{Prediction error from truth of a 3 satellite constellation using EvolveGCN without physics loss.}
    \label{fig:3sat_prediction}
\end{figure}

\begin{figure}[H]
    \centering
    \includegraphics[width=0.85\textwidth]{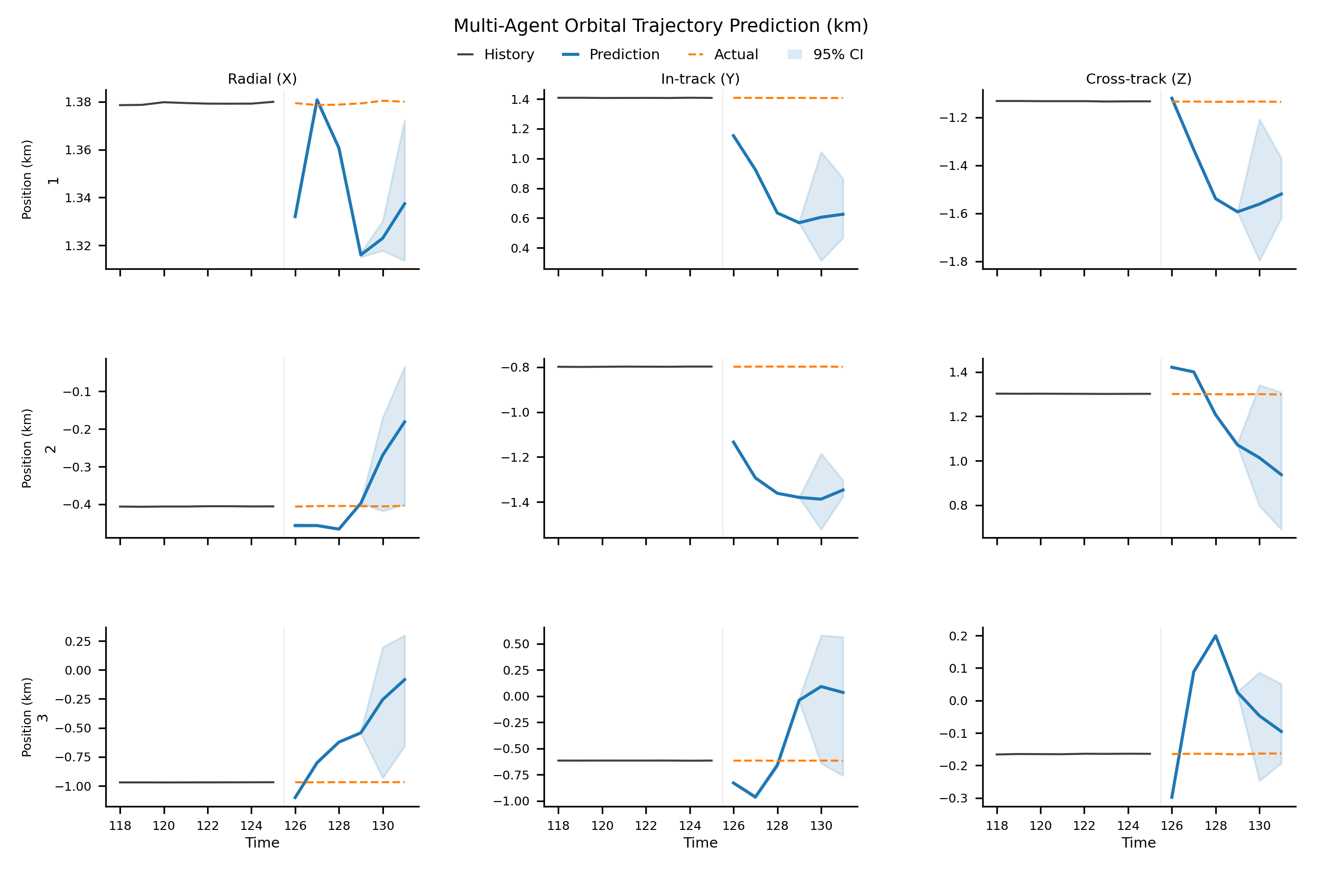}
    \caption{Prediction comparison of a 3 satellite constellation using EvolveGCN with physics loss.}
    \label{fig:3sat_prediction}
\end{figure}

\begin{figure}[H]
    \centering
    \includegraphics[width=0.85\textwidth]{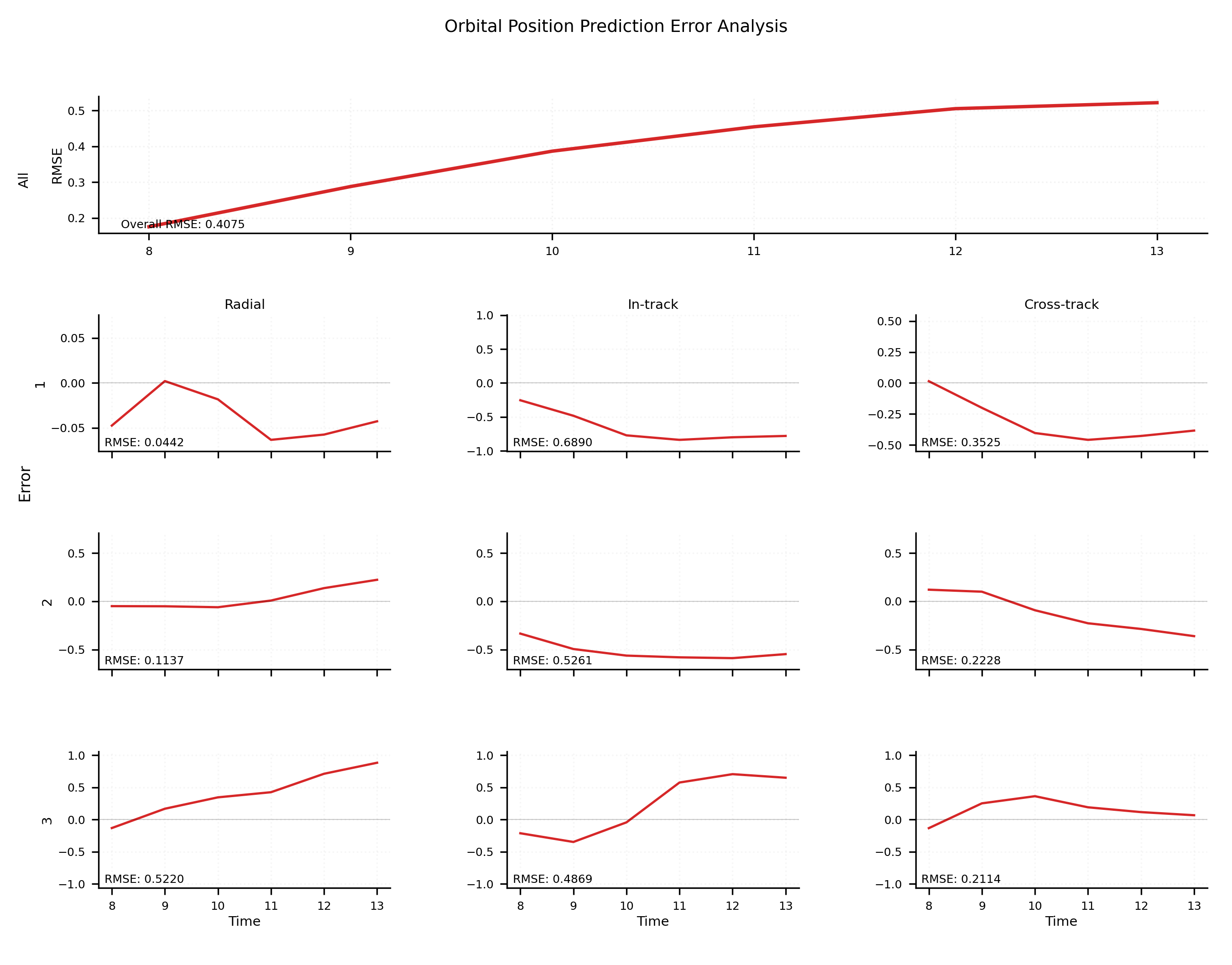}
    \caption{Prediction error from truth of a 3 satellite constellation using EvolveGCN with physics loss.}
    \label{fig:3sat_prediction}
\end{figure}

The preliminary results can be condensed into the table shown below. 

\begin{table}[h!]
\centering
\caption{RMSE Comparison Between EvolveGCN With and Without Physics Loss Across 3 Satellites}
\label{tab:physics_vs_nophysics}
\begin{tabular}{|c|c|c|c|}
\hline
\textbf{Satellite} & \textbf{Axis} & \textbf{RMSE (No Physics) [km]} & \textbf{RMSE (With Physics) [km]} \\
\hline
\multirow{3}{*}{Sat 1} 
  & In track     & 0.4895  & 0.6890 \\
  & Cross track  & 0.1421  & 0.3525 \\
  & Radial       & 0.1758  & 0.0442 \\
\hline
\multirow{3}{*}{Sat 2} 
  & In track     & 0.5076  & 0.5261 \\
  & Cross track  & 0.0363  & 0.2228 \\
  & Radial       & 0.1897  & 0.1137 \\
\hline
\multirow{3}{*}{Sat 3} 
  & In track     & 0.6392  & 0.4869 \\
  & Cross track  & 0.7169  & 0.2214 \\
  & Radial       & 0.4659  & 0.5220 \\
\hline
\end{tabular}
\end{table}

\vspace{12pt}

The results shown in Figures as well as summarized in Table 2 show that including the physics into the loss does show limited improvement in the prediction of some of the satellites orbital positions as well as improvements in the overall stability of the prediction over time. Table 1 shows the final error values after prediction and from the results, the cross track position is shown to be slightly more accurate across all satellites in the no physics loss test. Some fluctuation in the other coordinate parameters so marginal improvement with the physics loss. Investigating figures 2 and 4 show that the inclusion of physics loss lowers the RMSE of the system as a whole. Observing the trends of the total RMSE of the systems also shows that inclusion of physics allows for a steadier increase in RMSE towards the end of the prediction horizon rather than the sharp increase that no physics has. This error could be reduced with proper hyperparameter turning as both models where trained on the same model architecture the only change was one with physics loss included and other was not included. With proper hyperparameter tuning the errors in the physics informed EvovleGCN should be less and more stable over longer periods in the future. Another tuning parameter that should increase the accuracy of the physics results is the inclusion or removal of links in this satellite constellation. A fully connected graph, or one that every agent is connected to every other agent, may lead to more errors in the aggregation of data to predict future positions. The inverse can be said for a sparse graph where not enough data is being shared between agents leading to poor position estimation relative to agents. Finding the proper number of connections needed for this kind of prediction is critical. Both of these plots where using around 60 percent connection including self connections. This means that lowering this connection destiny value could allow for better results with the physics loss as this seems to improve the predictions of different axes positions based on the number of connections each agent has and to what agent neighbor. The physics informed network shows some interesting trends and could lead to very accurate results when tuned and trained on a diverse dataset. 

Changes from this test and the final results will be an optimized architecture that displays the absolute best results that can come from this model. This will be tested on a larger dataset consisting of thousands of trajectories with various number of agents. This diverse dataset will allow for a single model to be introduced to a system after training so each agent can predict its neighbors next movements in the future based on historical context. This will lead to future research papers exploring tactics to use this data to make a better informed decision on how to avoid and keep formations.

\section{Conclusion}
This extended abstract displays the capabilities with including physics in EvolveGCN to predict the potions of multiple agents using graphical representations and GCNs. This preliminary research results display stable predictions over a horizon using physics when compared to not using physics in the loss function for EvolveGCN. This research could lead to implementation of obstacle and collision avoidance for multi-agent systems utilizing graph theory and graph neural networks. 

\bibliography{refrences}

\begin{thebibliography}{26}
\newcommand{\enquote}[1]{``#1''}
\providecommand{\natexlab}[1]{#1}
\providecommand{\url}[1]{\texttt{#1}}
\providecommand{\urlprefix}{URL }
\expandafter\ifx\csname urlstyle\endcsname\relax
  \providecommand{\doi}[1]{\discretionary{}{}{}https://doi.org/#1}\else
  \providecommand{\doi}[1]{\discretionary{}{}{}\urlstyle{rm}\url{https://doi.org/#1}}\fi

\bibitem[{Brunke et~al.(2022)Brunke, Greeff, Hall, Yuan, Zhou, Panerati, and Schoellig}]{Brunke2022}
Brunke, L., Greeff, M., Hall, A.~W., Yuan, Z., Zhou, S., Panerati, J., and Schoellig, A.~P., \enquote{Safe Learning in Robotics: From Learning-Based Control to Safe Reinforcement Learning,} \emph{Annual Review of Control, Robotics, and Autonomous Systems}, Vol.~5, 2022, pp. 411--444.
\newblock \doi{10.1146/annurev-control-042920-020211}.

\bibitem[{Sheikhsamad and Puig(2024)}]{Sheikhsamad2024}
Sheikhsamad, M., and Puig, V., \enquote{Learning-Based Control of Autonomous Vehicles Using an Adaptive Neuro-Fuzzy Inference System and the Linear Matrix Inequality Approach,} \emph{Sensors}, Vol.~24, No.~8, 2024, p. 2551.
\newblock \doi{10.3390/s24082551}.

\bibitem[{Shalaby et~al.(2021)Shalaby, Champagne~Cossette, Forbes, and Le~Ny}]{Shalaby2021RAL}
Shalaby, M., Champagne~Cossette, C., Forbes, J.~R., and Le~Ny, J., \enquote{Relative Position Estimation in Multi-Agent Systems Using Attitude-Coupled Range Measurements,} \emph{IEEE Robotics and Automation Letters}, Vol.~6, No.~3, 2021, pp. 4955--4962.
\newblock \doi{10.1109/LRA.2021.3067253}.

\bibitem[{Wang et~al.(2023)Wang, Wang, Li, and Zhao}]{Wang2023Sensors}
Wang, S., Wang, Y., Li, D., and Zhao, Q., \enquote{Distributed Relative Localization Algorithms for Multi-Robot Networks: A Survey,} \emph{Sensors}, Vol.~23, No.~5, 2023, p. 2399.
\newblock \doi{10.3390/s23052399}.

\bibitem[{Chang et~al.(2025)Chang, Yang, Zhang, Jiao, Cheng, and Fu}]{Chang2025}
Chang, X., Yang, Y., Zhang, Z., Jiao, J., Cheng, H., and Fu, W., \enquote{Consensus-Based Formation Control for Heterogeneous Multi-Agent Systems in Complex Environments,} \emph{Drones}, Vol.~9, No.~3, 2025, p. 175.
\newblock \doi{10.3390/drones9030175}.

\bibitem[{Olfati-Saber et~al.(2007)Olfati-Saber, Fax, and Murray}]{olfati2007consensus}
Olfati-Saber, R., Fax, J.~A., and Murray, R.~M., \enquote{Consensus and cooperation in networked multi-agent systems,} \emph{Proceedings of the IEEE}, Vol.~95, No.~1, 2007, pp. 215--233.
\newblock \doi{10.1109/JPROC.2006.887293}.

\bibitem[{Ren et~al.(2005)Ren, Beard, and Atkins}]{ren2005survey}
Ren, W., Beard, R.~W., and Atkins, E.~M., \enquote{Survey of consensus problems in multi-agent coordination,} \emph{American Control Conference}, 2005, pp. 1859--1864.
\newblock \doi{10.1109/ACC.2005.1470307}.

\bibitem[{Jiang et~al.(2022)Jiang, Wang, and Guo}]{jiang2022multi}
Jiang, W., Wang, R., and Guo, C., \enquote{Multi-agent graph neural networks for collaborative perception and decision-making: A survey,} \emph{IEEE Transactions on Neural Networks and Learning Systems}, 2022.
\newblock \doi{10.1109/TNNLS.2022.3186842}.

\bibitem[{Zhou et~al.(2021)Zhou, Cui, Hu, Zhang, Yang, Liu, Wang, Li, and Sun}]{zhou2021graphneuralnetworksreview}
Zhou, J., Cui, G., Hu, S., Zhang, Z., Yang, C., Liu, Z., Wang, L., Li, C., and Sun, M., \enquote{Graph Neural Networks: A Review of Methods and Applications,} , 2021.
\newblock \urlprefix\url{https://arxiv.org/abs/1812.08434}.

\bibitem[{Pareja et~al.(2020)Pareja, Domeniconi, Chen, Ma, Suzumura, Kanezashi, Kaler, Schardl, and Leiserson}]{pareja2020evolvegcn}
Pareja, A., Domeniconi, G., Chen, J., Ma, T., Suzumura, T., Kanezashi, H., Kaler, T., Schardl, T., and Leiserson, C., \enquote{Evolvegcn: Evolving graph convolutional networks for dynamic graphs,} \emph{Proceedings of the AAAI conference on artificial intelligence}, Vol.~34, 2020, pp. 5363--5370.

\bibitem[{Hays et~al.(2024)Hays, Henderson, Miller, Phillips, and Soderlund}]{doi:10.2514/1.G008280}
Hays, C.~W., Henderson, T., Miller, K., Phillips, S., and Soderlund, A., \enquote{Angles-Only Cooperative Local Catalog Maintenance of Close-Proximity Satellite Systems,} \emph{Journal of Guidance, Control, and Dynamics}, Vol.~47, No.~12, 2024, pp. 2573--2586.
\newblock \doi{10.2514/1.G008280}, \urlprefix\url{https://doi.org/10.2514/1.G008280}.

\bibitem[{Hays et~al.()Hays, Miller, Soderlund, Phillips, and Henderson}]{doi:10.2514/6.2024-0992}
Hays, C.~W., Miller, K., Soderlund, A.~A., Phillips, S., and Henderson, T., \emph{State Omniscience for Cooperative Local Catalog Maintenance of Close Proximity Satellite Systems}, ????
\newblock \doi{10.2514/6.2024-0992}, \urlprefix\url{https://arc.aiaa.org/doi/abs/10.2514/6.2024-0992}.

\bibitem[{Valencia-Palomo et~al.(2024)Valencia-Palomo, Targui, and Lopez-Estrada}]{ValenciaPalomo2024}
Valencia-Palomo, G., Targui, B., and Lopez-Estrada, F.-R., \enquote{Consensus Tracking Control of Multiple Unmanned Aerial Vehicles Subject to Distinct Unknown Delays,} \emph{Machines}, Vol.~12, No.~5, 2024, p. 337.
\newblock \doi{10.3390/machines12050337}.

\bibitem[{Yan et~al.(2024)Yan, Zhou, and Yang}]{yan2024control}
Yan, L., Zhou, J., and Yang, K., \enquote{Control-Aware Trajectory Predictions for Communication-Efficient Drone Swarm Coordination in Cluttered Environments,} \emph{arXiv preprint arXiv:2401.12852}, 2024.

\bibitem[{Yan et~al.(2018)Yan, Xiong, and Lin}]{yan2018stgcn}
Yan, S., Xiong, Y., and Lin, D., \enquote{Spatial temporal graph convolutional networks for skeleton-based action recognition,} \emph{AAAI Conference on Artificial Intelligence}, Vol.~32, 2018.

\bibitem[{Zhao et~al.(2020)Zhao, Zhang et~al.}]{zhao2020stgcn}
Zhao, L., Zhang, J., et~al., \enquote{ST-GCN-based trajectory prediction in crowd navigation,} \emph{Pattern Recognition Letters}, 2020.

\bibitem[{Kim et~al.(2021)Kim, Kim, and Han}]{kim2021stgrnn}
Kim, H., Kim, Y.~M., and Han, S., \enquote{ST-GRNN: Spatio-temporal graph recurrent neural network for traffic prediction,} \emph{IEEE Transactions on Intelligent Transportation Systems}, 2021.
\newblock \doi{10.1109/TITS.2021.3061404}.

\bibitem[{Manessi et~al.(2020)Manessi, Rozza, and Manzo}]{manessi2020dynamic}
Manessi, F., Rozza, A., and Manzo, M., \enquote{Dynamic graph convolutional networks,} \emph{Pattern Recognition}, Vol.~97, 2020, p. 107000.

\bibitem[{Zhou et~al.(2022)Zhou, Wang, and Ma}]{zhou2022dgnn}
Zhou, J., Wang, L., and Ma, C., \enquote{Multi-agent path planning via dynamic graph attention networks,} \emph{IEEE/RSJ International Conference on Intelligent Robots and Systems (IROS)}, 2022.

\bibitem[{Rossi et~al.(2020)Rossi, Chambers, Frasca, Eynard, Monti, and Bronstein}]{rossi2020temporal}
Rossi, E., Chambers, B., Frasca, F., Eynard, D., Monti, F., and Bronstein, M.~M., \enquote{Temporal graph networks for deep learning on dynamic graphs,} \emph{arXiv preprint arXiv:2006.10637}, 2020.

\bibitem[{Han et~al.(2022)Han, Lin, and Zhang}]{han2022tgn}
Han, Y., Lin, F., and Zhang, W., \enquote{TGN-based multi-agent interaction modeling for trajectory prediction,} \emph{IEEE Transactions on Intelligent Transportation Systems}, 2022.

\bibitem[{Li et~al.(2023)Li, Wang, and Chen}]{li2023tgn}
Li, X., Wang, Q., and Chen, Y., \enquote{Event-driven spatiotemporal modeling for UAV swarms using temporal graph neural networks,} \emph{Sensors}, 2023.

\bibitem[{Raissi et~al.(2019)Raissi, Perdikaris, and Karniadakis}]{raissi2019physics}
Raissi, M., Perdikaris, P., and Karniadakis, G.~E., \enquote{Physics-informed neural networks: A deep learning framework for solving forward and inverse problems involving nonlinear partial differential equations,} \emph{Journal of Computational Physics}, Vol. 378, 2019, pp. 686--707.

\bibitem[{Karniadakis et~al.(2021)Karniadakis, Kevrekidis, Lu, Perdikaris, Wang, and Yang}]{karniadakis2021physics}
Karniadakis, G.~E., Kevrekidis, I.~G., Lu, L., Perdikaris, P., Wang, S., and Yang, L., \enquote{Physics-informed machine learning,} \emph{Nature Reviews Physics}, Vol.~3, No.~6, 2021, pp. 422--440.

\bibitem[{Li et~al.(2022)Li, Zhang, Liu, and Wang}]{li2022deepopinn}
Li, Z., Zhang, Y., Liu, M., and Wang, P., \enquote{DeepOPINN: Data-driven operator learning for solving parametric PDEs via physics-informed neural networks,} \emph{Computer Methods in Applied Mechanics and Engineering}, Vol. 398, 2022, p. 115248.

\bibitem[{Lu et~al.(2021)Lu, Kroemer, and Schaal}]{lu2021learning}
Lu, F., Kroemer, O., and Schaal, S., \enquote{Learning Contact-Rich Manipulation Skills with Guided Policy Search and Physics-Based Losses,} \emph{Robotics: Science and Systems (RSS)}, 2021.

\end{thebibliography}

\end{document}